\documentclass[showpacs,showpacs,aps,twocolumn]{revtex4}

\usepackage{graphicx}
\usepackage{dcolumn}
\usepackage{bm}

\begin{document}

\title{Delocalization and spin-wave dynamics in ferromagnetic chains
with long-range correlated random exchange}

\author{F.~A.~B.~F. de Moura, M.~D. Coutinho-Filho,
E.~P. Raposo}

\affiliation{Laborat\'orio de F\'{\i}sica Te\'orica e Computacional,
Departamento de F\'{\i}sica, Universidade Federal de Pernambuco,
50670-901 Recife, PE, Brazil}

\author{M.~L. Lyra}

\affiliation{Departamento de F\'{\i}sica, Universidade Federal de Alagoas
Br 101 Km 14, Cidade Universit\'{a}ria, Macei\'o, AL, 57072-970. Brazil}

\begin{abstract}

We study the one-dimensional quantum Heisenberg ferromagnet with exchange
couplings exhibiting long-range correlated disorder
with power spectrum proportional to $1/k^{\alpha}$, where $k$
is the wave-vector of the modulations on the random
coupling landscape. By using renormalization group, integration of the
equations of motion and exact diagonalization, we compute
the spin-wave localization length and the mean-square displacement
of the wave-packet.  We find that, associated with the emergence of extended
spin-waves in the low-energy region for $\alpha > 1$, the wave-packet
mean-square displacement changes from a long-time
super-diffusive behavior for $\alpha <1$ to a
long-time ballistic behavior for $\alpha > 1$. At the vicinity of $\alpha =1$,
the mobility edge separating the  extended and localized phases is
shown to scale with the degree of correlation as $E_c\propto (\alpha -1)^{1/3}$.
\end{abstract}

\pacs{75.10.Jm, 75.30.Ds, 75.50.Lk}

\maketitle

\section{Introduction}

It is well established that  one-electron  eigenstates in
chains with uncorrelated disorder are exponentially localized~\cite{abrahams}.
However, several theoretical investigations have shown that a series of
one-dimensional versions of the Anderson Model  exhibit a breakdown
of the Anderson's localization induced by internal short-range
 correlations on the disorder distribution, including hopping and on-site energy correlations~\cite{flores1}
or just on-site energy correlations~\cite{dunlap,evangelou2}.
Recently, it  has been demonstrated that the one-dimensional Anderson model with
long-range correlated disorder can
display a phase of extended electronic states~\cite{chico,izrailev}.
These results have been
confirmed by  microwave transmission spectra of  single-mode
waveguides with inserted correlated scatterers~\cite{apl2}.
The influence of long-range disorder on the electron motion
in two dimensions has been recently investigated~\cite{efetov} and  a
two-dimensional  layered Anderson model with long-range correlated
disorder has been shown to exhibit a Kosterlitz-Thouless-type
metal-insulator transition~\cite{liu}.

It is also known that the magnon equation of motion for
ferromagnetic spin chains with uncorrelated random  nearest-neighbor
exchange couplings can be exactly mapped onto  an electronic chain
with a particular form of off-diagonal disorder where the random
hopping integrals appear correlated in pairs~\cite{ziman,evangelou1}.
By using a perturbation approach combined with a scaling hypothesis it was
demonstrated~\cite{ziman} that the  singularities of the density of states and the localization
length of a random ferromagnetic Heisenberg chain depend on the distribution of
exchange couplings. For uncorrelated random exchange couplings $J\in[0,1]$ with a
probability distribution function $P(J)=(1 - \delta) J^{-\delta}$,
it was demonstrated that for $\delta <-1$,  which implies $<1/J>$ and $<1/J^2>$ finite, the density
of states $\rho (E)$  diverges as $1/E^{1/2}$ and the localization length as $1/E$.
Distinct regimes emerge when either of the above two moments diverges, thus  generalizing
other studies~\cite{theodorou}. Moreover, in order  to elucidate doubts raised in the
literature~\cite{marek}, the spin-wave dynamics of the one-dimensional Heisenberg ferromagnet
was investigated~\cite{evangelou1} for the above power-law distribution of exchange couplings $J$.
The mean-square displacement of the wavepacket asymptotically  displays super-diffusion
dynamics [$\sigma(t)^2 \propto t^{3/2}$]
for weak disorder ($\delta < 0$), diffusive behavior [$\sigma(t)^2 \propto t^{1}$]
for $\delta = 0$, and localization [$\sigma(t)^2 \propto \mbox{constant}$]
for strong disorder ($\delta > 0$). Therefore, the uncorrelated Heisenberg ferromagnet
with random exchange couplings can display a super-diffusive dynamics if $<1/J>$ is finite.
In all three cases, a ballistic regime [$\sigma(t) ^2 \propto t^{2}$] emerges for
initial times. By using the transfer matrices technique,  the
singularities of the density of states and localization length were verified.
The super-diffusive behavior is closely related to the one found in the random-dimer
version of the Anderson model~\cite{dunlap}.
~

In  this paper, we study the nature of the spin-wave modes of a quantum Heisenberg
ferromagnetic chain with long-range correlated random exchange couplings
$J_n$ assumed to have  spectral
power density $S \propto 1/k^{\alpha}$. A previous study reported
some finite-size
scaling evidences of the emergence of a phase of extended low-energy
excitations~\cite{delson}.
Here, we will use a renormalization group technique to provide accurate estimates for
 the mobility-edge energy as a function of the degree of correlation and  to obtain
 the scaling behavior governing the vanishing of the mobility edge in the vicinity of
 $\alpha=1$.
Further, we also study the quantum diffusion of the wave-packets in these chains using
direct integration of the motion equations and exact diagonalization to investigate the possible
emergence of a new dynamical regime associated with the occurrence of extended low-energy spin
waves for $\alpha >1$.

\section{The Model and Renormalization-Group Calculation}

We consider a Hamiltonian model describing a
spin$-1/2$ quantum ferromagnetic Heisenberg chain of $N$ sites with random exchange
couplings $J_n$:
\begin{equation}
H=\sum _{n=-N/2}^{N/2} J_n {\bf S_n} \cdot {\bf S_{n+1}},
\end{equation}
where ${\bf S_n}$ represents the quantum spin operator at site $n$ and open
boundary conditions are used.
We take the exchange couplings $J_n$ connecting sites $n$ and $n+1$
to be  correlated in such a sequency to describe the trace of a fractional
Brownian motion~\cite{feder,osborne,greis}:
\begin{equation}
J_n = \sum_{k=1}^{N/2}
 \left[k^{-\alpha }\left(\frac{2\pi }{N}\right)^{(1-\alpha )}\right]^{1/2}
\cos{\left( \frac{2\pi nk}{N} +\phi_k\right)},
\end{equation}
where $k$ is the wave-vector of the modulations on the random
coupling landscape and $\phi_k$ are $N/2$ random phases uniformly distributed
in the interval $[0,2 \pi ]$. The exponent $\alpha$ is directly related to the
Hurst exponent $H$ ($\alpha =2H+1$) of the rescaled range analysis, which describes
the self-similar character of the series and the
persistent character of its increments: for $\alpha >2$ ($H>1/2$) the increments
are persistent, while for $\alpha <2$ ($H<1/2$) they are
anti-persistent. In the case of $\alpha=2$ ($H=1/2$)  the sequency of exchange
couplings resembles the trace of the usual Brownian motion, while for
$\alpha=0$ ($H=-1/2$) one recovers the uncorrelated random exchange Heisenberg model.
The coupling distribution is Gaussian for $\alpha =0$ but assumes a non-Gaussian
form
for finite $\alpha$ once the presence of long-range correlations implies
in the lack of self-averaging and the breakdown of the central limit theorem. In order to avoid a vanishing
exchange coupling we shift all couplings generated by Eq.~(2) such to have average value $<J_n>=-4.5$
and  variance $<\Delta J_n>=1$. Note that, in such case, all moments of
the resulting distribution are finite. In order to keep the variance size independent,
the normalization factor scales with the chain size. A detailed
finite-size scaling analysis has shown how such normalization
procedure is reflected in the main character of the one-magnon
excitations~\cite{delson}. Indeed, without such rescaling of the potential the disorder
width diverges for any $\alpha\neq 0$ and all states are expected to remain localized~\cite{russ}.
~

The ground state of the system contains all spins pointing in the same direction.
If a spin deviation occurs
at a site $n$, this excited state  is described by
\begin{equation}
\phi_n =S_n^{+}|0>,
\end{equation}
where the operator $S_{n}^{+}$ creates a spin deviation at site $n$ and $ |0>$ denotes
 the ground state. The eigenstates of the  Hamiltonian are therefore a
linear combination of $\phi_n$, i.e., $\Phi=\sum_n c_n \phi_n$,  the coefficients
$c_n$ satisfying  the equation below~\cite{evangelou1,theodorou,marek}
\begin{equation}
(J_{n-1}+J_n)c_n - J_{n-1}c_{n-1} - J_nc_{n+1} = 2Ec_n,
\end{equation}
where $E$ is the excitation energy.
~

In order to study the properties of one-magnon states, we apply
a decimation \linebreak renomalization-group technique. The method is based on the particular form
assumed by the equation of motion satisfied by the Green's operator matrix elements
$[G(E)]_{i,j}=\langle i|1/(E-H)|j\rangle$\cite{chao,chico}:
\begin{eqnarray}
(E-\epsilon _{n+\mu}^{0})[G(E)]_{n+\mu,n}=\delta_{\mu,0} +
J_{n+\mu,n+\mu-1}^0[G(E)]_{n+\mu-1,j} +
J_{n+\mu,n+\mu+1}^0[G(E)]_{n+\mu+1,n},
\end{eqnarray}
where $\epsilon _n^0=(J_{n-1}+J_n)/2$ and $J_{n,n+1}^0=J_{n+1,n}^0 = J_n/2$.
After eliminating the matrix  elements associated with a given site,
the remaining set of equations of motion can be expressed in the same form
as the original one, but with renormalized parameters:
\begin{eqnarray}
\epsilon_{N}^{(N-1)}(E)&=&\epsilon_{N}+J_{N-1,N}\frac{1}{E-\epsilon_{N-1}^{(N-2)}(E)}
J_{N-1,N},\\
J_{0,N}^{(eff)}(E)&=&J_{0,N-1}^{(eff)}\frac{1}{E-\epsilon_{N-1}^{(N-2)}(E)}J_{N-1,N} ,
\end{eqnarray}
where, after $N-1$ decimations, $\epsilon_{N}^{N-1}$ denotes the renormalized diagonal element at site
$N$ and $J_{0,N}^{(eff)}$ indicates the effective renormalized
exchange coupling connecting the sites $0$ and $N$.
~

To investigate the localized/delocalized nature of the spin-wave modes, we
compute
the inverse of the  excitation width or the Lyapunov coefficient $\gamma(E)$ (inverse localization length).
 The Lyapunov coefficient is
 asymptotically related to the effective exchange coupling by~\cite{chao,chico}:
\begin{equation}
\gamma(E)=-\lim_{N \rightarrow \infty}[\frac{1}{N}\ln|J_{0,N}^{(eff)}(E)|] .
\end{equation}
After a linear regression of $\ln{|J_{0,N}^{(eff)} (E)|}$  versus $N$ we have a
 direct extrapolation
of the  Lyapunov coefficient in the thermodynamical limit.
We  computed $\gamma(E)$ for distinct values of
the exponent $\alpha$ and $N = 10^5$ sites. In addition, the density of states (DOS) was
calculated by using the numerical Dean's method~\cite{dean}. In Fig.~1 we show
the normalized DOS for chains with $N=10^5$ sites. Notice that it becomes
less rough as $\alpha$ is increased and its  singularity  at the bottom
of the band is not affected by the imposed long-range correlation
in the coupling constants~(see insets). For $\alpha=1.5$  it consists of a non-fuctuating part near
the bottom of the band with the same form as that of the pure chain~($J_n=\mbox{constant}$).
Previous studies have pointed out that the smoothing of the DOS is usually connected with the
emergence of delocalized states~\cite{alts}.

In Fig.~2 we show the plot of
$\gamma$ versus $E$  for $\alpha =0$ (uncorrelated random exchanges).
In this case, the Lyapunov coefficient
is finite for all energies, except
for $E=0$ as usual. This behavior remains qualitatively unaltered for $0 < \alpha \le 1$,
therefore implying in the absence of extended spin waves in this regime.
For $\alpha = 0$, we have also observed that, near the bottom of the band, the Lyapunov coefficient vanishes
as $\gamma \propto E^\nu$
with $\nu =1.0$, in agreement with Ref.~\cite{ziman} for probability distribution
functions
with finite $<1/J>$ and $<1/J^2>$~(see inset of Fig.~2).
The picture is qualitatively different for $\alpha >1$.  In Fig.~3(a) we plot
 $\gamma$ versus $E$ for $\alpha =1.5$. The Lyapunov coefficient
vanishes within a finite range of energy values, thus confirming the presence
of low-energy extended spin waves. In all chains studied with sizes ranging from $10^5$ up
to  $10^6$ sites,
the  $\gamma(E)$ curves appear to be the same, indicating
that the extended phase of magnons is stable in the thermodynamical limit.
The phase diagram in the ($E_c$, $\alpha$) plane is shown in
Fig.~3(b), with $E_c$~(given in units of $\Delta J$)
denoting the mobility edge and statistical errors are smaller than the symbol sizes.
The data analysis (see inset) suggests that, at the vicinity of $\alpha =1$, the mobility edge depends
on the correlation exponent as $E_c \propto (\alpha-1)^{\gamma}$, with $\gamma=1/3$.

\section{Spin-Wave Dynamics}

In order to investigate  the spin-wave dynamics, we compute the time dependence of
the mean-square displacement of the wave-packet. Let us consider an excitation
initially localized at site $n_0$, represented at $t=0$ by its eigenfunction
$\phi _n(t=0) =\delta _{n,n_0}$. Its  time evolution is described by the
Schr\"odinger equation ($\hbar =1$):
\begin{equation}
i\frac{d\phi _n(t)}{dt}=H\phi _n(t),
\end{equation}
whose  time-dependent wave-function can be written in terms of the computed eigenvectors
$V^{(j)}$ and eigenvalues $E_j$ of $H$ as~\cite{evangelou1}
\begin{equation}
\phi _n(t)= \sum_j^N V_n^{(j)} V_{n_0}^{(j)}\exp (-iE_jt).
\end{equation}
The same time-dependent wave-function can be
obtained by integrating the equations of motion.
The second moment of the corresponding spatial probability distribution is then given by
\begin{equation}
\sigma ^2(t)=\sum _n (n-n_0)^2\phi _n(t)\phi _n^{*}(t) .
\end{equation}

From the mean-square displacement $\sigma^2(t)$ we can estimate the wave-packet
spread in space at a time $t$.
For any $\alpha \ge 0$, we find ballistic behavior [$\sigma(t) ^2 \propto t^{2}$]
for initial times, indicating that the disorder have not yet been realized by
the spin waves~\cite{evangelou1}. In the case of  uncorrelated random exchange ($\alpha=0$)
the self-expanded chain was used to minimize end effects. When the probability of finding
the particle at the ends of the chain exceeded $10^{-100}$ we added new
sites thus expanding the chain. In Fig.~4 we show the mean squared displacement
versus time for $\alpha=0$ as obtained by integrating the equations of motion.
In this case, for longer times the wave-packet presents a super-diffusive spread
[$\sigma(t) ^2 \propto t^{3/2}$]  in agreement with previous studies~\cite{evangelou1}
for uncorrelated random exchange distribution with $<1/J>$ finite. In Fig.~5 we
plot the data obtained by  integrating the equations of motion for $\alpha=0.5$
and $20000$ sites. As indicated, the initial ballistic motion extends over
longer times, although a super-diffusive motion still takes place after this
initial transient. Finally, Fig.~6 shows that for $\alpha=1.5$  the wave-packet
displays only a ballistic spread. In this case our
calculation was performed by using both numerical integration of the
equations of motion for $N=20000$ sites and exact diagonalization for $N=2000$,
$4000$ and $8000$ sites, in which case the end effect is present~(see inset).

In contrast with the case of uncorrelated random exchange
couplings~\cite{evangelou1}, in which an asymptotic super-diffusive
behavior sets up  whenever $<1/J>$ is finite, we find a crossover
with increasing $\alpha$ from super-diffusive to ballistic asymptotic
regimes induced by long-range correlations in the exchange couplings. The ballistic
regime for $\alpha>1$ can be understood following arguments
similar to those used in Ref.~\cite{dunlap,evangelou1}. Exploring the exact mapping of
the magnon problem onto a paired electronic one,  the diffusion
coefficient $D$ can be estimated by integrating $v(k)\lambda(k)$ over the
extended states that effectively participate in the transport,
where $v(k)$ and $\lambda(k)$ are respectively the velocity and
mean free-path of the excitation mode with wavenumber $k$. In a
finite chain all extended modes have $\lambda(k)\simeq N$ and
travel with finite velocity since in the electronic problem the DOS is
nonsingular near the band center. Once there is a finite fraction of  states
that are delocalized, the integration runs over a finite wavenumber and,
interchanging $N$ and $t$,  the diffusion coefficient results $D\propto t$.
Consequently, the mean square displacement
$\langle\sigma^2(t)\rangle = Dt\propto t^2$ confirming the
ballistic nature of the wave-packet spread found in our numerical
analysis.

\section{Conclusions}

In summary, we have studied the one-dimensional quantum Heisenberg
ferromagnet with exchange couplings  exhibiting long-range
correlated disorder with spectral power density proportional to
$1/k^{\alpha}$. By using a decimation renormalization-group
technique we have found further evidences suggesting that
this magnetic system displays a phase of extended spin waves in the
low-energy region for $\alpha >1$~($H>0$). The mobility edge
separating low-energy extended and high-energy localized states
was
shown to depend on the degree of correlation in a very
special manner. Finally, through integration of the equations of motion
and exact diagonalization, we have also computed the
mean-square displacement of the spin-wave packet.  For $0 < \alpha \le 1$,
we have found long-time super-diffusion, in agreement with previous
works for uncorrelated random exchange distribution with $<1/J>$
finite. However, for strong correlations ($\alpha > 1$) a long-time ballistic  regime
was
numerically observed which is associated with the emergence of extended excitations.
We believe that the reported results might be
useful to stimulate further theoretical and experimental
investigations of spin-wave dynamics on correlated ferromagnetic
chains and non-periodic ferromagnetic super-lattices.

This work was supported by the Brazilian agencies CNPq, Finep, CAPES, FACEPE, and FAPEAL.

\newpage

\noindent
{\bf \Large Figure Captions}

\normalsize

\vspace*{0.6cm}
\noindent
{\bf Fig.~1}~The spin-wave density of states (DOS) for chains with $N=10^5$ sites
using Dean's method. The DOS becomes less rough as $\alpha$ is increased and its
singularity  at the bottom of the band is not affected by the imposed
long-range correlation in the coupling constants~(insets).
For $\alpha=1.5$  it consists of a non-fuctuating part near the
bottom of the band with the same form as that of the pure chain~($J_n=\mbox{constant}$).

\vspace*{0.3cm}
\noindent
{\bf Fig.~2}~Lyapunov coefficient $\gamma$ versus energy $E$ for $\alpha = 0$
(uncorrelated  random exchange model) and $N=10^5$ sites from the
renormalization procedure. The Lyapunov coefficient is finite for non-zero energies (localized states),
and vanishes as $\gamma \propto E^\nu$, $E \rightarrow 0$, with $\nu =1.0$ (inset).

\vspace*{0.3cm}
\noindent
{\bf Fig.~3}~(a)~Lyapunov coefficient $\gamma$ versus
energy $E$ for $\alpha = 1.5$ and $N=10^5$ sites from the renormalization procedure. The Lyapunov
coefficient vanishes within a finite range of energy values revealing the presence
of extended low-energy spin waves. (b)~$(E_c, \alpha)$ phase diagram, where $E_c$ is the mobility edge
(in units of $\Delta J$) for $N = 10^5$ sites. The phase of extended
spin-waves emerges for $\alpha > 1$ and $E < E_c$.

\vspace*{0.3cm}
\noindent
{\bf Fig.~4}~Mean squared displacement $\sigma^2$ versus time $t$ for $\alpha = 0$ from the
integration of the equations of motion. The self-expanded chain was used to minimize end effects.
The spread  of the wavepacket depicts a crossover from a initial ballistic
($\sigma^2 \propto t^2$) to a super-diffusive ($\sigma^2 \propto t^{3/2}$) behavior.

\vspace*{0.3cm}
\noindent
{\bf Fig.~5}~Mean squared displacement $\sigma^2$ versus time $t$  for $\alpha = 0.5$
and $N=20000$ sites from the integration of the equations of motion . A longer living ballistic
motion ($\sigma^2 \propto t^{2}$) is found but still followed by a crossover to
the super-diffusive regime ($\sigma^2\propto t^{3/2}$).

\vspace*{0.3cm}
\noindent
{\bf Fig.~6}~Mean squared displacement $\sigma^2$ versus time $t$  for $\alpha = 1.5$ for $N=20000$
sites from the integration of the equations of motion. Inset: results using exact
diagonalization for $N=2000$~(dotted line), $4000$~(dashed line)  and $8000$~(dot-dashed line) sites.
Ballistic behavior ($\sigma^2 \propto t^{2}$) is found for all times.

\end{document}